\documentclass[aps,prb,twocolumn,showpacs,groupedaddress]{revtex4-1}

\usepackage{graphicx}
\usepackage{dcolumn}
\usepackage{bm}
\usepackage{amsmath}
\usepackage{natbib}
\usepackage{lineno}
 \usepackage{array}

\begin{document}

\preprint{APS/123-QED}

\title{Surface acoustic wave driven ferromagnetic resonance in (Ga,Mn)(As,P) epilayers}

\author{L. Thevenard$^{1,2}$\email[e-mail: ]{thevenard@insp.jussieu.fr}, C. Gourdon$^{1,2}$, J. Y. Prieur$^{1,2}$, H. J. von Bardeleben$^{1,2}$, S. Vincent$^{1,2}$, L. Becerra$^{1,2}$,  L. Largeau$^{3}$, and J.-Y. Duquesne$^{1,2}$}

 
\affiliation{
$^1$ CNRS, UMR7588, Institut des Nanosciences de Paris, 4 place Jussieu,75252 Paris, France\\
$^2$ Sorbonne Universit\'es, UPMC Univ Paris 06, UMR7588 4 place Jussieu,75252 Paris, France
$^3$  Laboratoire de Photonique et Nanostructures, CNRS, UPR 20, Route de Nozay, Marcoussis, 91460, France\\
}

\date{\today}

\label{sec:Abstract}

\begin{abstract}


Interdigitated transducers were used to generate and detect surface acoustic waves on a thin layer of (Ga,Mn)(As,P). The out-of-plane uniaxial magnetic anisotropy of this dilute magnetic semiconductor is very sensitive to the strain of the layer, making it an ideal test material for the dynamic control of magnetization via magneto-striction. The time-domain measurement of the amplitude and phase of the transmitted SAW during magnetic field sweeps indicated a clear resonant behavior at a field close to the one calculated to give a precession frequency equal to the  SAW frequency. A resonance was observed from 5K to 85K, just below the Curie temperature of the layer. A full analytical treatment of the coupled magnetization/acoustic dynamics showed that the magneto-strictive coupling modifies the elastic constants of the material and accordingly the wave-vector solution to the elastic wave equation. The shape and position of the resonance were well reproduced by the calculations, in particular the fact that  velocity (phase) variations resonated at lower fields than the acoustic attenuation variations.
\end{abstract}


\maketitle

\section{Introduction}


Magnetostriction is the interaction between strain and magnetization, which leads to a change in a magnetic sample's shape when its magnetization is modified\cite{Lacheisserie2004}. The opposite effect, inverse magnetostriction, whereby magnetization can be changed upon application of a strain, is particularly relevant to magnetic data storage technologies as a possible route towards induction-free data manipulation when used dynamically. It has  been proposed for magnetization switching through resonant \cite{Thevenard2013} or non-resonant processes  \cite{Li2012,Davis2010}. Experimental results on magneto-acoustic coupling have  been obtained in various configurations, either all optical\cite{Scherbakov2010} or all-electric\cite{Pomerantz1961,bommel59,Feng1982,Weiler2011,Dreher12,Polzikova2013}. The latter consists in the radio-frequency (rf) excitation of interdigitated transducers (IDTs) deposited onto a piezoelectric/ferromagnetic bilayer. This has for instance led to the extensive theoretical and experimental investigation  of magnetization precession triggered by surface or bulk acoustic waves (resp. SAWs, BAWs) in Ni based films since the 1960s\cite{Pomerantz1961,bommel59,Ganguly1976,Feng1982,Dreher12}. Elegant data has also been obtained more recently on Yttrium Iron Garnet, where BAWs were used to build a magnetic field tunable acoustic resonator\cite{Polzikova2013}.

 No  demonstration of SAW-induced ferromagnetic resonance has been shown in dilute magnetic semi-conductors (DMS). Yet these materials, such as  (Ga,Mn)(As,P), are an interesting system since their magnetostrictive coefficients  vary strongly with temperature. This provides an excellent tool to develop and validate theoretical models. Their low Curie temperature (100 - 180~K) imposes to generate SAWs at cryogenic temperatures, but their magnetization precession frequencies  are  close to accessible SAW frequencies (GHz) and their small and tunable magnetic anisotropy make them a good candidate for SAW-assisted magnetization switching\cite{Thevenard2013}.


In this paper, we evidence experimentally SAW-driven ferromagnetic resonance in a thin film of (Ga,Mn)(As,P) excited  at 549~MHz, between 5 and 85~K (Sec. \ref{exp}). Both acoustic attenuation and velocity variations are monitored in the time-domain. We then solve the coupled magnetization and elastic dynamics equations and determine with a good match to  the experimental data (Sec. \ref{analysis}) the expected resonance fields and acoustic resonance shape.

\section{Experimental methods}
\label{exp}

 \begin{figure}
	\centering
	\includegraphics[width=0.45\textwidth]{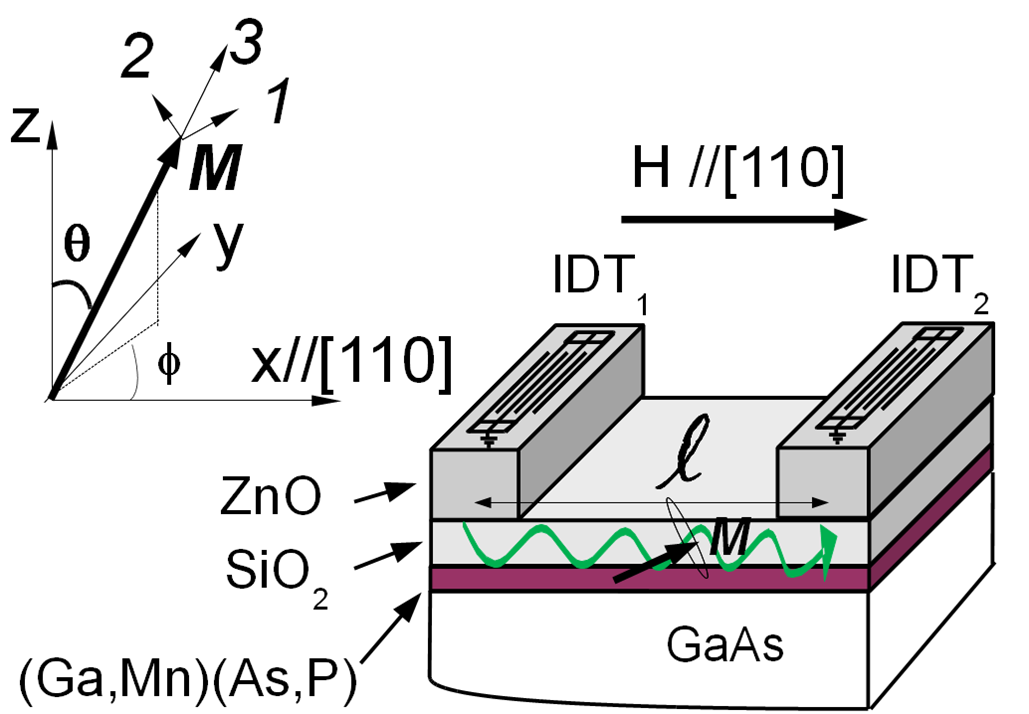}
	\caption{Structure of the sample (not to scale). 50 nm ferromagnetic epilayer, 70~nm SiO$_{2}$ buffer, 250 nm piezoelectric ZnO. The  IDTs are separated by 2~mm, but the effective length of the delay line is taken center-to-center of the IDTs, i.e $l$=2.3~mm. (upper left) Definition of the (x,y,z) and (1,2,3) reference frames.}
	\label{fig:Fig1}
\end{figure}

A $d$=50~nm thick layer of (Ga$_{1-x}$,Mn$_{x}$)(As$_{1-y}$,P$_{y}$) was grown  by molecular beam epitaxy. After a 1h/250$^{\circ}$C anneal, its Curie temperature reached T$_{c}$ = 105~K and its active Mn concentration $x^{eff}\approx 3.5\%$. Since GaAs is  only weakly piezoelectric, a 70/250~nm bilayer of SiO$_{2}$/ZnO was sputtered onto the magnetic layer. The silica underlayer was required for good adhesion. Care was taken to keep the substrate holder at relatively low temperature (150$^\circ$C) during the ZnO deposition so as to not further anneal the magnetic layer. The Phosphorus ($y \approx 4\%$) was necessary to induce  tensile strain in the layer, in order to obtain a dominantly uniaxial magnetic anisotropy\cite{lemaitre08,Cubukcu2010a}, spontaneously aligning the magnetization perpendicular-to-plane. The resulting lattice mismatch of the layer to the substrate was of -1610~ppm. 

\begin{figure}
	\centering
	\includegraphics[width=0.5\textwidth]{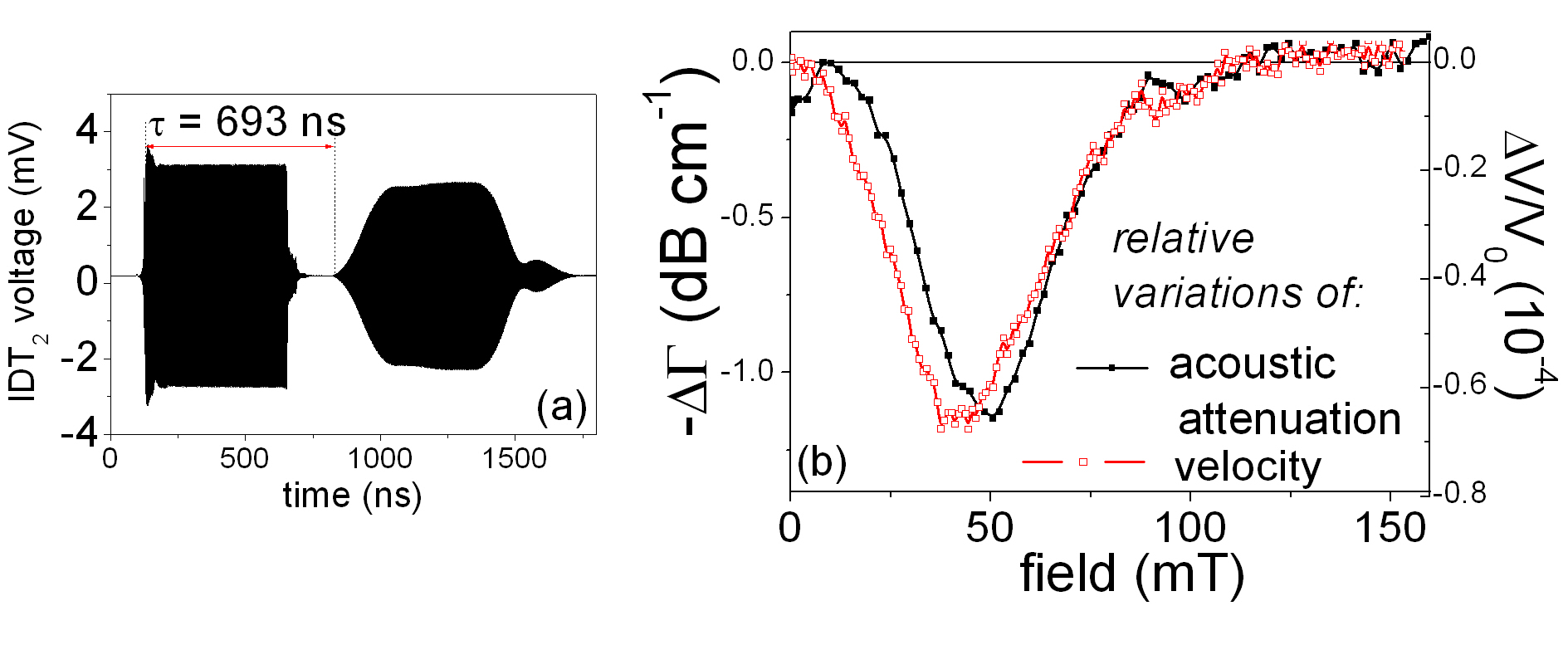}
	\caption{(a) Receiving IDT signal: the electromagnetic field radiated by the emitter is shortly followed by the transmitted surface acoustic wave. T=120K, 549 MHz. (b) Attenuation changes and relative velocity variations at  T~=~80K. The opposite of $\Delta \Gamma$ has been plotted in order to highlight the different resonant field from the velocity variations.}
	\label{fig:Fig2}
\end{figure}

Cr/Au interdigitated transducers (sixty pairs, thickness 10/80~nm) were then evaporated and a window  opened in the ZnO layer between the two IDTs (Fig. \ref{fig:Fig1}). The metallization ratio of 0.5 and teeth width of 1.25~$\mu$m, yielded an acoustic wavelength of 5~$\mu$m. The emitter (IDT$_{1}$) was excited by 550~ns bursts of rf voltage modulated at 1~kHz. After propagation along the [110] direction, the SAW was detected by the receiver IDT$_{2}$  and the signal was acquired with a digital oscilloscope over typically 4000 averages.  This time-domain technique allowed us to  (i) verify that the SAWs were indeed generated/detected in the sample, and (ii) clearly separate the antenna-like radiation of IDT$_{1}$ (traveling at the speed of light), from the acoustic echo (traveling at the Rayleigh velocity), as shown in Fig. \ref{fig:Fig2}a.  The transit time lies around $\tau$=693~ns, which immediately gives an experimental estimation of the Rayleigh velocity V$_r\approx$~2886 m~s$^{-1}$. The transfer function of the device exhibited the typical band-pass behavior centered at the 549 MHz resonance frequency (FWHM of 8 MHz). The power applied to the IDT$_{1}$ was of +20dBm (100 mW) on a 50 ohm load. The excitation frequency   was $\omega/2\pi$=549~MHz.

The field/temperature-dependent measurements were done in a cryostat allowing rf cables to be brought down to the sample, which could be cooled down to 2.7 K.  Unless specified, the field was applied  in the plane of the sample, along the SAW wave-vector, i.e along a hard magnetic axis, and swept at 0.23~mT~s$^{-1}$. A phase detection scheme then yielded the amplitude $A$ and the phase $\phi = \omega\tau$ of the transmitted SAW. The phase variations $\Delta \phi$ were converted into relative velocity variations using $\Delta V /V_{0}$ = $\frac{\Delta \phi}{\omega\tau_{0}}$. The attenuation changes were computed using $\Delta \Gamma$ = -$\frac{20}{l}\log \frac{A}{A_{0}}$. $A_{0}$ is an arbitrary reference amplitude. $l=2.3 mm$ and $\tau_{0}$ = 797ns are the IDTs~' center-to-center distance and the corresponding transit time.


\section{Experimental results}
\label{exp}

\begin{figure}
	\centering
	\includegraphics[width=0.35\textwidth]{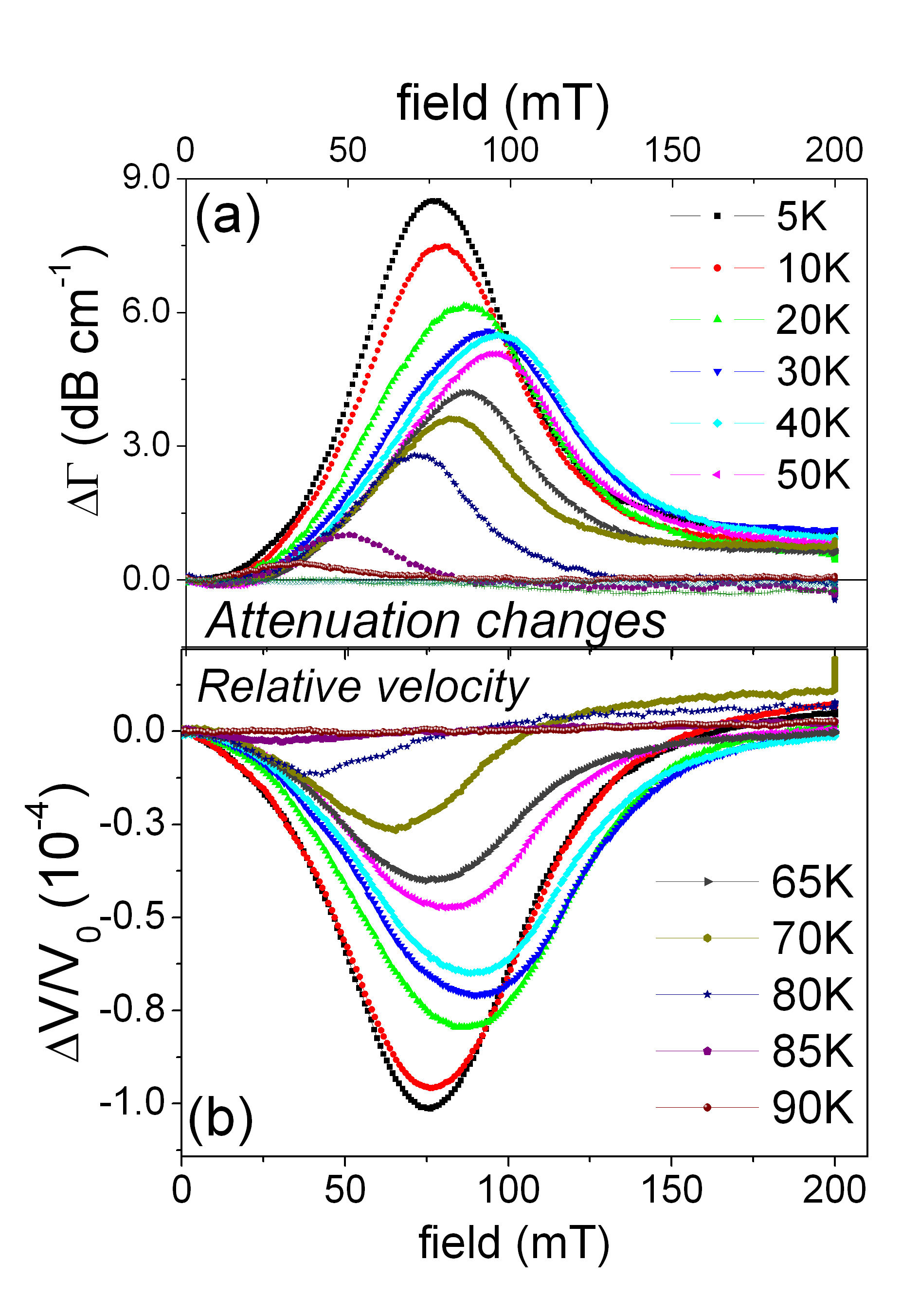}
	\caption{(a) Variation of acoustic attenuation and (b) relative velocity change of the SAW between 5~K and 90~K.  }
	\label{fig:Fig3}
\end{figure}

A typical sweep at T~=~ 80K is shown in Fig. \ref{fig:Fig2}b. Acoustic attenuation and velocity variations were both identical at low and high fields, but showed a clear feature at a particular field, hereafter called resonance field. The resonance disappeared above 90~K. Measurements down to 5~K showed that the amplitude of the effect steadily increased with decreasing temperature (Fig. \ref{fig:Fig3}). The resonance field was however not monotonous with temperature, lying within 35-94~mT with a maximum at 30-40~K. The resonance width followed the same trend, within the bounds 9-17~mT (error bars in Fig. \ref{fig:Fig4}b). All curves shared the following features: a fairly symmetrical, non-hysteretic resonance, with the velocity variations  systematically resonating at a lower field than the amplitude variations.  The maximum variation of acoustic attenuation, $\Delta \Gamma$=~8.5~dB~cm$^{-1}$ was observed at $T$=5~K. It remains weak compared to the value of 200 dB~cm$^{-1}$ measured at 2.24 GHz on a similar device on Nickel\cite{Weiler2011}. This is due to both the higher SAW frequency used by these authors, as the amplitude variations are directly proportional to $\omega$ (see Eqs. (24,27) of Ref. \onlinecite {Dreher12} for instance), and the much lower magneto-strictive constants found in DMS. These are defined as the fractional change in sample length as the magnetization increases from zero to its saturation value and their maximum values lie around $|\lambda_{100}|\approx$ ~ 9~ppm for (Ga,Mn)As\cite{Masmanidis2005} and  $\lambda_{100}\approx$50~ppm for Nickel\cite{Lacheisserie2004}. Finally, we can easily check that the resonance frequency of the transducers is only very weakly modified by the magneto-elastic interaction: since $|\Delta \phi/\phi_{0}|$=$|\Delta \omega/\omega|$$\approx10^{-4}$, typical velocity variations measured at $\omega/2\pi$=549~MHz yield a frequency change of a negligible $\approx$55~kHz.

\section{Model}
\label{analysis}

 \begin{figure}
	\centering
	\includegraphics[width=0.45\textwidth]{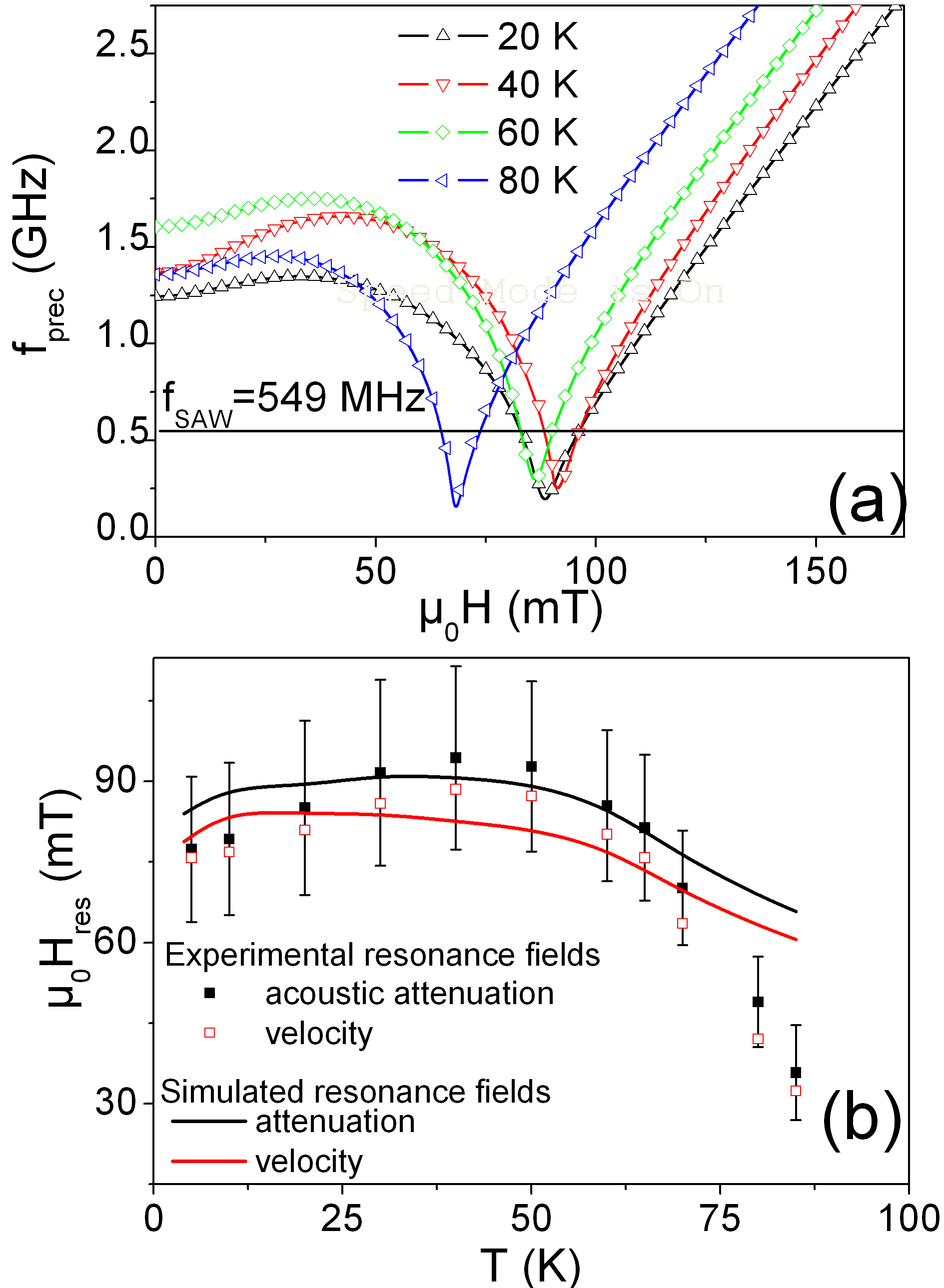}
	\caption{(a) Calculated precession frequency  versus field applied along [110], no sample tilt.  The horizontal line indicates the SAW frequency. (b) Measured (symbols) and simulated resonance fields (dashed line, sample tilt 1.2$^{\circ}$, taking into account both $A_{2\varepsilon}$  and $A_{4\varepsilon}$)   versus temperature for the attenuation (black) and velocity (red) variations. The error bars correspond to the full width at half maximum of the resonance.}
	\label{fig:Fig4}
\end{figure}

We have shown above that at a particular applied field, the transmitted SAW was slightly absorbed (by a 19$\%$ decrease in amplitude at 5 K), and delayed (by about...90 ps !) through its  interaction with the magnetization of the (Ga,Mn)(As,P) layer.  To confirm that this is indeed SAW-driven ferromagnetic resonance, we calculate the expected resonance fields and shapes. Microscopically, the resonance may be seen as the crossing of  magnon and phonon dispersion curves at the wave-vector imposed by the IDTs, $k_{SAW}$. Macroscopically, the total energy of the system may then be written\cite{Landau1984} :

\begin{align}
\label{eq:Etot}
E_{tot} & = W+M_{s}F_{mc}+M_{s}F_{ms}   
	\end{align}
	
	With: 
	
\begin{align}
\label{eq:W}
W & =\frac{1}{2}c_{ijkl}\varepsilon_{ij}\varepsilon_{kl}=W_{0}+W_{SAW}(t)\\
	\label{eq:Fms}
F_{ms} & = F_{ms,0}+F_{ms,SAW}\\
			 & = (\varepsilon_{zz}-\frac{\varepsilon_{xx}+\varepsilon_{yy}}{2})[(A_{2\varepsilon}+A_{4\varepsilon})m_{z}^{2}\\
\nonumber   & +\frac{A_{4\varepsilon}}{2}m_{z}^{4}+A_{4\varepsilon}(m_{x}^{4}+m_{y}^{4})]\\
F_{mc} & = -\mu_{0}\vec{H}.\vec{m}+[\frac{\mu_0M_s}{2}-3B_{c}]m_{z}^{2}+\frac{5}{2}B_{c}m_{z}^{4}\\
\nonumber   &-B_{c}(m_{x}^{4}+m_{y}^{4})+\frac{B_{2||}}{4}(m_{x}^{2}-m_{y}^{2})
\end{align}



The components of the unit  magnetization vector are defined as $m_{i}$=$M_{i}/M_{s}$  (i=x,y,z)  where $M_{s}$ is the magnetization at saturation and x//[110]. $W$ is the purely elastic contribution and $F_{mc}$ is the purely magnetic energy (magneto-crystalline, demagnetizing and Zeeman contributions, in units of field). In the latter, $H$ is the applied field, $B_{c}$ the cubic  anisotropy constant and $B_{2||}$  the uniaxial one, distinguishing  in-plane [110] and [1-10] axes\cite{B2par}.  $F_{ms}$ is the magneto-elastic contribution (in units of field) where the magnetostrictive coefficients $A_{2\varepsilon}$, $A_{4\varepsilon}$    depend on both the static  strain felt by the layer ($\varepsilon_{xx,0},\varepsilon_{yy,0},\varepsilon_{zz,0}$), and the dynamic SAW-induced strain ($\varepsilon_{xx,SAW}(t)$, $\varepsilon_{zz,SAW}(t)$). The $\varepsilon_{xz,SAW}(t)$ component of the SAW does not have any magnetostrictive action on the layer, and $\varepsilon_{yy,SAW}(t)$ is  not excited by our set-up. The total strain components are thus given by $\varepsilon_{ii}=\varepsilon_{ii,0}+\varepsilon_{ii,SAW}(t)$.  The magnetization and acoustics dynamics are then obtained by solving the  Landau-Lifshitz Gilbert equation (LLG, Eq. 6) and  the elastic wave equation (EWE, Eq. 7): 


\begin{align}
   \label{eq:LLG}
\frac{\partial \vec{m}}{\partial t} & = \frac{\gamma}{M_s} \vec{m} \times\vec{\nabla}_{\vec{m}} E_{tot}+\alpha \vec{m} \times \frac{\partial{\vec{m}}}{\partial t}\\
	\label{eq:EWE}
\rho\frac{\partial^{2}R_{tot,i}}{\partial t^{2}} & = \frac{\partial\sigma_{ik}}{\partial x_{k}}=\frac{\partial}{\partial x_{k}}\frac{\partial E_{tot}}{\partial\varepsilon_{ik}}
\end{align}

where $\alpha$ is an effective damping constant and $\gamma$ the gyromagnetic factor.  $\vec{m}$=$ \vec{m_0}+\vec{m}(t)$ is the sum of the equilibrium  magnetization unit vector and the rf magnetization and likewise for the displacement $\vec{R}_{tot}$=$\vec{R}_{0}+\vec{u}(t)$.  The displacements are related to the strain by $\varepsilon_{ij}=\frac{\partial R_{tot,i}}{\partial x_{j}}$, $\rho$ is the material density and  $c_{ijkl}$ the elastic constant tensor  defined in the x,y,z frame (see Appendix B). Note that, as assumed by other authors\cite{Ganguly1976,Dreher12} the exchange contribution was neglected in Eq. (6), as the typical SAW wave-vector ($\approx 1/\Lambda_{SAW}$) is much smaller than the first spin-wave wave-vector ($\approx 1/d$), leading to an essentially flat magnon dispersion curve for the frequencies considered here. For this reason, although we should in all rigor be talking about "spin-wave FMR", we will use the shorter term "ferromagnetic resonance".

Following Dreher et al.\cite{Dreher12}, we define a second reference frame (1,2,3) where $\vec{m_{3}}$ is aligned with the static magnetization (polar coordinates ($\theta_0$, $\phi_0$), see Appendix A). We are then left with two sets of unknowns: $(m_{1},m_{2})(t)$ (magnetization dynamics) and  $(u_{x},u_{z})$(t) (acoustic dynamics), as the transverse displacement $u_{y}$ cannot be excited by our device. Solving Eq. (6) in the linear approximation with $m_i(x,t)$=$m_{0,i}e^{i(\Omega t-kx)}$ leads to the following system:

\begin{equation}
	\label{detm}
\left(\begin{array}{c}
m_{1}\\
m_{2}
\end{array}\right)=[\chi]
\left(\begin{array}{c}
\mu_{0}h_{1}\\
\mu_{0}h_{2}
\end{array}\right)
\end{equation}

The dynamic fields are defined by $\mu_{0}h_{i}=-\frac{\partial F_{ms,SAW}}{\partial m_{i}}\mid_{\vec{m}=\vec{m}_{3}}$. Since the uniaxial term $A_{2\varepsilon}$ ($\approx$ 40-60T) is around 10 times larger than $A_{4\varepsilon}$, the  $A_{4\varepsilon}$ terms will be neglected. The dynamic magneto-elastic energy then simply reads: $F_{ms,SAW}=A_{2\varepsilon}\Delta\varepsilon(t)m_{z}^{2}$,  so that $\mu_{0}h_{1}$=$A_{2\varepsilon}\Delta \varepsilon(t)(cos^{2}\theta_{0}-sin2\theta_{0}m_{1})$ and $\mu_{0}h_{2}$=0.

 The susceptibility tensor $[\chi]$ (given in Appendix A)  depends on the static magnetic anisotropy constants, the damping and the SAW excitation frequency $\omega$. Canceling the determinant of $[\chi]^{-1}$  yields the precession frequency (real part of $\Omega$)  $\left(\frac{\omega_{prec}}{\gamma}\right)^{2}=(F_{11}-F_{33})(F_{22}-F_{33})-F_{12}^{2}$ where the terms $F_{ij}$ stand for $\frac{\partial^2 (F_{mc}+F_{ms,0})}{\partial m_i \partial m_j}$. Fig. \ref{fig:Fig4}a shows the field-dependence of this precession frequency at various temperatures, calculated from the FMR anisotropy coefficients. $f_{prec}(\mu_0 H)$ first decreases, crossing the  SAW frequency of 549 MHz (full line in Fig. \ref{fig:Fig4}a) at a particular field. When the magnetization is aligned with the field (saturated), $f_{prec}$ reaches a minimum. After saturation, the resonance frequency increases with field, and crosses $f_{SAW}$ a second time. We will show below that this second crossing does not give rise to any magneto-acoustic resonance. The cross-over fields of  $f_{prec}(\mu_0 H)$ with  $f_{SAW}$ (Fig. \ref{fig:Fig4}a) can already give a good approximation of the expected resonance fields.  It is however not sufficient to explain why the  resonance fields are different for relative variations of the SAW attenuation and velocity. For this it is necessary to calculate how the SAW wave-vector is modified by its interaction with the (Ga,Mn)(As,P) layer.


We place ourselves in the semi-infinite medium approximation and assume the SiO$_2$ layer to be a small perturbation to the system since its thickness is much smaller than $\Lambda_{SAW}$ (see Appendix C for details on this point). With a general form of displacement, $u_{i}(x,t)$=$U_{i}e^{-\beta z}exp[i(\omega t-kx)]$ and using the equilibrium conditions on the strain, the EWE  may then be simplified into:

\begin{eqnarray}
\label{det1solution}
\left(\rho\omega^{2}+(\frac{A_{\chi}}{4}-c_{11})k^{2}+c_{44}\beta^{2}\right)u_x+(c_{44}+c_{13}+\frac{A_{\chi}}{2})\beta ik u_z & =0\\
(c_{44}+c_{13}+\frac{A_{\chi}}{2})\beta ik u_x+\left(\rho\omega^{2}+(c_{33}-A_{\chi})\beta^{2}-k^{2}c_{44}\right)u_z & =0
\end{eqnarray}

Here we have introduced the complex constant:

\begin{equation}
	A_{\chi}=M_{s}A_{2\varepsilon}^{2}sin^{2}(2\theta_{0})\chi_{11}\\
	\label{Achi}
\end{equation}

Two  features come out. Firstly, this system is the formal equivalent of the solution to the  EWE in a cubic, non-magnetostrictive material, with three of the elastic constants modified as follows: 

\begin{align}
\label{det}
\begin{array}{c}
c_{13}\mapsto c_{13}^{\prime}=c_{13}+A_{\chi}/2\\
c_{11}\mapsto c_{11}^{\prime}=c_{11}-A_{\chi}/4\\
c_{33}\mapsto c_{33}^{\prime}=c_{33}-A_{\chi}
\end{array}
\end{align}

The elastic constants are modified through $A_{\chi}$ which depends on the applied field, the anisotropy constants and the SAW frequency (through $\chi_{11}$). The real part of 	$A_{\chi}$ represents at most $\approx$10$\%$ of the GaAs elastic constants.  This constant embodies the physics of the coupled magnon-phonon system as it modifies the elastic constants of the material. It cancels out when the material ceases to be magneto-strictive ($A_{2\varepsilon}$=0) and/or when the magnetization is colinear or normal to the SAW wave-vector. This is why no acoustic resonance is observed at the second crossing of $f_{prec}(\mu_0 H)$ with  $f_{SAW}$, once the magnetization is aligned with the applied field ($	A_{\chi}|_{\theta_0=\pi/2}=0$). To check this point, we repeated the experiment with the field applied perpendicular to the sample this time : no resonance was observed, either in the attenuation changes or in the velocity variations.

Secondly, using the full depth-dependence of the displacements results in a coupling of the $u_{x}$ and $u_{z}$ components ($\beta$ terms in Eqs. 9,10), contrary to  simpler cases treated previously\cite{Dreher12}. In fact, it is through the $z$ attenuation that $c_{13}$ and $c_{33}$ constants are modified by the magneto-strictive interaction; they would otherwise be left unchanged. 
 
 Canceling the determinant of Eqs. (9,10) yields two solutions with the corresponding absorption coefficients $\beta_{1,2}$ and  x,z  amplitude ratios $U_{z}/U_{x}$=$r_i$ (i=1,2, see Appendix D). As neither of these  satisfy the normal boundary condition $\sigma_{xz}|_{z=0}=0$ at the vacuum interface, a linear combination of these two solutions needs to be considered:

\begin{eqnarray}
	u_{x}=\left[U_{x1}\exp(-\beta_{1}z)+U_{x2}\exp(-\beta_{2}z)\right]e^{i(\omega t-kx)}\\
	u_{z}=\left[U_{z1}\exp(-\beta_{1}z)+U_{z2}\exp(-\beta_{2}z)\right]e^{i(\omega t-kx)}
\label{déplacements CL}
\end{eqnarray}

The boundary conditions $\sigma_{xz}|_{z=0}=\sigma_{zz}|_{z=0}$=0 now lead to a new system, similar to Eqs. (9,10). Replacing  $r_i$, $\beta_i$ by their expressions and using $\omega/V_r$=$k$, its determinant  eventually leads to  Eq. \eqref{det2}. This implicit polynomial  equation in $k$ may be solved  numerically to yield the wave-vector solutions  $k_{sol}$ in presence of magneto-strictive interaction. There are three distinct physical solutions to Eq. \eqref{det2}, but only the Rayleigh surface wave can be excited by our device\cite{BAWs}.

In the absence of magneto-striction, the usual Rayleigh velocity\cite{Cottam73} $V_r=\frac{\omega}{k_{sol}|_{A_{\chi}}=0}$=2852.2 m~s$^{-1}$ is recovered, very close to the crude experimental estimation made earlier.

	\begin{widetext}
	\begin{align}
\label{det2}
\left(c_{44}-\rho \frac{\omega^2}{k^2}\right)\left[c_{11}^{\prime}c_{33}^{\prime}-c_{13}^{\prime2}-c_{33}^{\prime}\rho \frac{\omega^2}{k^2}\right]^{2} &= c_{33}^{\prime}c_{44}\left(c_{11}^{\prime}-\rho \frac{\omega^2}{k^2}\right)\left(\rho \frac{\omega^2}{k^2}\right)^{2}
	\end{align} 
	\end{widetext}

The  amplitude of the transmitted SAW wave-vector is proportional to $\exp[-Im(k_{sol})l]$, and its phase is equal to $Re(k_{sol})l$. The relative variations are calculated with respect to the zero-field values.  We can now plot the expected relative variations of acoustic attenuation and velocity (e.g. at 40~K, Fig. \ref{fig:Fig6}) assuming we excite the IDTs at 549~MHz. In this calculation, we have also taken into account the $A_{4\varepsilon}$ term. The procedure is identical to the one described above, but the  expressions somewhat more cumbersome (see Appendix E for the corresponding effective elastic constants).

\section{Discussion}

\begin{figure}
	\centering
	\includegraphics[width=0.5\textwidth]{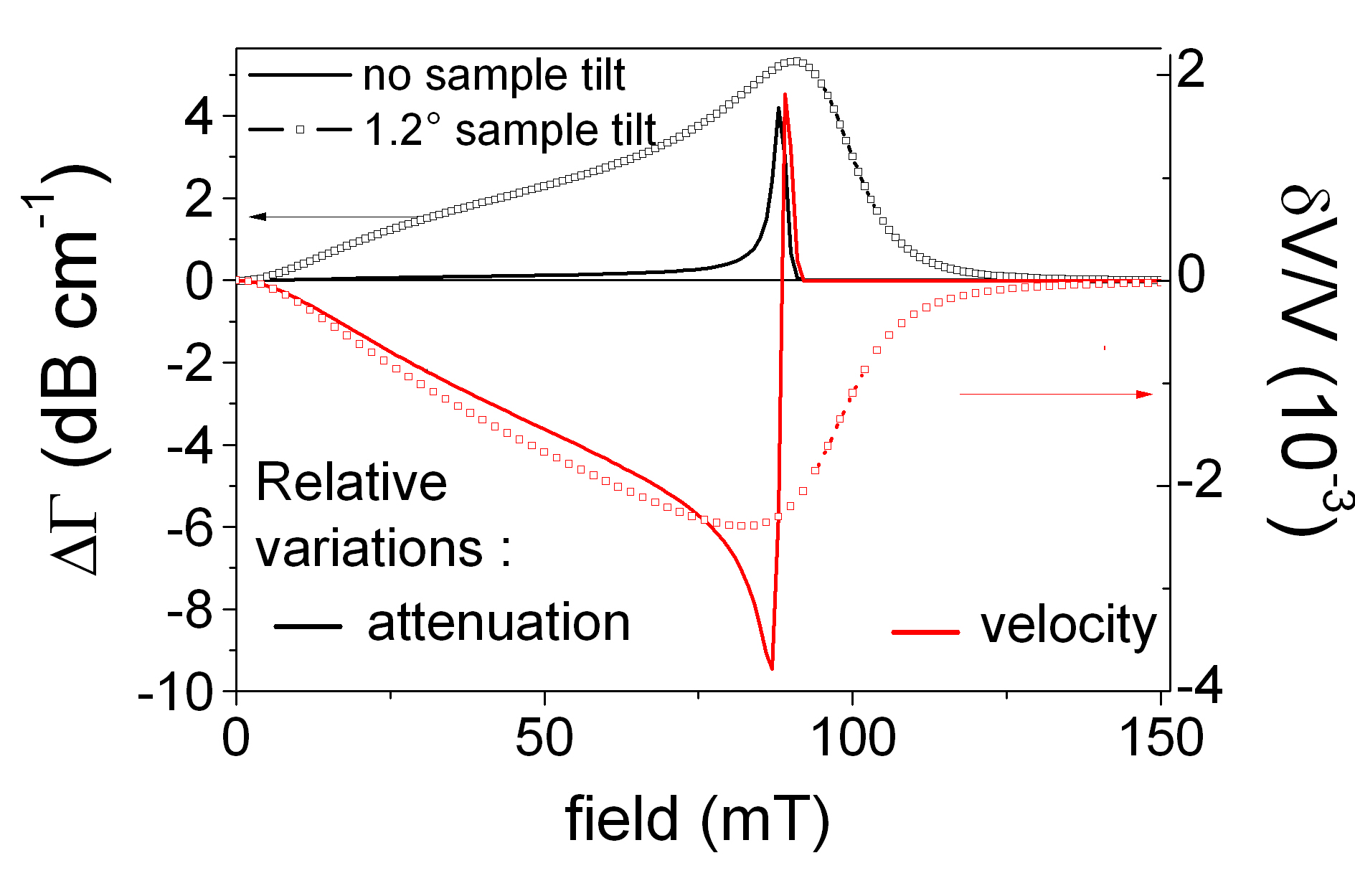}
	\caption{Relative variations of the acoustic attenuation (black) and velocity (red) calculated with the T=40 K micromagnetic parameters, $\alpha$=0.1 and $F$=0.105. The simulations were done taking into account both the $A_{2\varepsilon}$  and $A_{4\varepsilon}$ contributions, with (or without) a sample tilt in the (x,z) plane -  symbols (full lines). The relative  changes of attenuation and velocity without sample tilt have been divided by 20 for better visibility.}
	\label{fig:Fig6}
\end{figure}

The relative variation of \textit{attenuation}  (Fig. \ref{fig:Fig6}, full black line) is monopolar and peaks at 88~mT, as expected from the simple crossing of $f_{prec}(\mu_0$H) with $f_{SAW}=\omega/2\pi$ (Fig. \ref{fig:Fig4}). The relative variation of \textit{velocity}  (full red line) is bipolar, and cancels out when the amplitude variation is maximum. Both curves are quite asymmetric, plummeting to zero when the magnetization is aligned with the field (92~mT). Introducing a small 1.2$^{\circ}$ sample tilt in the (x,z) plane with respect to the field direction pushes the saturation field away from the resonance field, restoring the symmetry of the resonance. This tilt may have been introduced when gluing the sample. It  strongly reduces the magnitude of the effect, almost by a factor of 20. The attenuation resonance fields thus obtained are slightly higher than without tilt. The  higher-field bump of the velocity variations disappears, making the resonance unipolar and at lower fields than the amplitude variations, as observed experimentally. It is interesting to compare these results to those of Dreher et al.\cite{Dreher12}, computed using a similar approach for an in-plane Nickel thin film. Their closest comparable configuration is the one where the field is applied close to the sample normal (hard axis configuration). Their simulations (last line of Fig. 8 in Ref. \onlinecite{Dreher12}) also show that a bipolar shape is expected for the relative velocity, as the sample is excited closer and closer to its resonance frequency. Their experiments however also seem to show more of a monopolar behavior, for fields close to the sample normal.


 Simulated attenuation and velocity variations resonance fields are now plotted along with the experimental ones in Fig. \ref{fig:Fig4}b as a function of temperature. Their values are  well reproduced, so is their non-monotonous temperature variation. The latter can be traced back to a sign inversion of the $B_{2||}$ term with temperature i.e. a swap between [110] and [1-10] easy axes around 40~K.

Note that the magnetostrictive constants had to be  reduced by a filling factor $F$ to best reproduce the  amplitude of the effect since the magnetic layer occupies a small portion of the volume swept by the SAW: $A_{2\varepsilon}\mapsto FA_{2\varepsilon}$, $A_{4\varepsilon}\mapsto FA_{4\varepsilon}$. This effective medium approximation is routinely used in other solid state physics systems, such as the case of sparse quantum dots embedded in a wave-guide\cite{Melet2008}. A naive approach would lead us to expect $F\approx d/\Lambda_{SAW}$=$0.01$, whereas we converged to a value over ten times larger, $F=0.10$ to obtain a good agreement between simulated and experimental attenuation variations. The simulated velocity variations are then however off by about an order of magnitude compared to the experiment (Fig. \ref{fig:Fig6}). We believe however that this filling factor has little physical meaning.   Firstly, we have shown that not only $F$, but also the sample tilt play a great role in the amplitude of the effect, and this value is not known experimentally. Secondly,  the SAW amplitude is in fact not uniform across the depth $\Lambda_{SAW}$: it decreases rapidly away from the surface (see for instance Fig. 1b of Ref. \onlinecite{Thevenard2013}), so that the relative "weight" of the first $d$=50~nm is larger than $d/\Lambda_{SAW}$. To best reproduce quantitatively and qualitatively the shape and amplitude of the effect a more complete multi-layer approach using a transfer matrix formalism would clearly need to be adopted, as was for instance done in Ref. \onlinecite{Ganguly1976}.


\section{Conclusion}
\label{sec:Conclusion}


We have demonstrated the resonant excitation of magnetization precession in (Ga,Mn)(As,P) by a surface acoustic wave. Temperature-dependent measurements have clearly shown that the magnitude of the effect and the position of the resonance fields evolved with the magneto-strictive coefficient $A_{2\varepsilon}$. An analytical description of a SAW traveling through a magnetostrictive cubic medium  was derived, and the SAW dynamics was found to be given by  elastic tensor coefficients modified by a complex value proportional to $A^2_{2\varepsilon}$. This approach provides a very accurate way of predicting the resonance fields of SAW-driven FMR, an indispensable step towards SAW-induced precessional magnetization switching.

\label{sec:ACKNOWLEDGEMENTS}

This work was performed in the framework of the MANGAS and the SPINSAW projects (ANR 2010-BLANC-0424-02, ANR 13-JS04-0001-01). We thank A. Lema\^itre (Laboratoire de Photonique et Nanostructures) for providing the (Ga,Mn)(As,P) sample, and M. Bernard (INSP) for helping us with the cryogenic set-up.

\section*{APPENDIX}

\subsection{Magnetization dynamics}

Following Dreher et al.\cite{Dreher12}, the (1,2,3) reference frame is defined by $\vec{m}_3$ being aligned with the magnetization equilibrium position ($\theta_0$,$\phi_0$) and the following correspondence:

\begin{align}
\begin{array}{c}
m_{x}=m_{1}cos\theta_{0}cos\varphi_{0}-m_{2}sin\varphi_{0}+m_{3}sin\theta_{0}cos\varphi_{0}\\
m_{y}=m_{1}cos\theta_{0}sin\varphi_{0}+m_{2}cos\varphi_{0}+m_{3}sin\theta_{0}sin\varphi_{0}\\
m_{z}=-m_{1}sin\theta_{0}+m_{3}cos\theta_{0}
\end{array}
\end{align}

The susceptibility tensor defined in Eq. \eqref{detm} is given by:

\begin{align}
[\chi] =\frac{1}{D}\left(\begin{array}{cc}
F_{22}-F_{33}+\frac{i\alpha\omega}{\gamma} & -(F_{12}-\frac{i\omega}{\gamma})\\
-(F_{12}+\frac{i\omega}{\gamma}) & F_{11}-F_{33}+\frac{i\alpha\omega}{\gamma}
\end{array}\right) 
\end{align}

where $F_{ij}$=$\frac{\partial^2 (F_{mc}+F_{ms,0})}{\partial m_i \partial m_j}$ and

	\[D = (F_{11}-F_{33}+\frac{i\alpha\omega}{\gamma})(F_{22}-F_{33}+\frac{i\alpha\omega}{\gamma})-F_{12}^{^{2}}-\left(\frac{\omega}{\gamma}\right)^{2}
\]\\

  \subsection{Elastic coefficient tensor}

   The elastic coefficient tensor being defined in the reference frame of a cubic material, we must rotate it by $\pi$/4 for the particular case of a SAW traveling along [110]. The equivalence with the usual elastic constants\cite{Cottam73} $C^{0}_{ij}$ is: 
\\
 
 \begin{align}
\left\{ \begin{array}{c}
c_{11}=\frac{1}{2}C_{11}^{0}+\frac{1}{2}C_{12}^{0}+C_{44}^{0}\\
c_{12}=\frac{1}{2}C_{11}^{0}+\frac{1}{2}C_{12}^{0}-C_{44}^{0}\\
c_{13}=C_{12}^{0}\\
c_{33}=C_{11}^{0}\\
c_{44}=C_{44}^{0}\\
c_{66}=\frac{1}{2}C_{11}^{0}-\frac{1}{2}C_{12}^{0}
\end{array}\right.
 \end{align}
 
 Temperature variations of the elastic tensor have been neglected and (Ga,Mn)(As,P) elastic constants were assumed equal to those of GaAs. 
 
 \subsection{Influence of the SiO$_2$/ZnO on the (Ga,Mn)(As,P) static strain}

 An important question is whether the high temperature (150$^\circ$C) deposition of the SiO$_2$/ZnO bilayer modifies the magnetic layer's static strain. To check this, we performed room temperature high resolution x-ray rocking curves around the (004) reflection at different steps of the bilayer deposition. The lattice mismatch of the reference (unpatterned)  (Ga,Mn)(As,P) layer was around -1520~ppm, i.e under tensile strain on GaAs. After the SiO$_2$ deposition, the lattice mismatch dropped to -1360~ppm. However, the lattice  mismatch of the layer after  deposition of the full SiO$_2$/ZnO stack returned close to the reference value, around -1610~ppm, and remained unchanged after removal of the ZnO layer.  The rocking curves also pointed to the presence of a strain gradient extending into the GaAs substrate subsequently to the SiO$_2$/ZnO deposition. Given that the static strain of the magnetic layer seems to be affected by SiO$_2$/ZnO deposition, the FMR measurements of the magnetic anisotropy constants were done on the (Ga,Mn)(As,P)/SiO$_2$/ZnO stack after removal of the ZnO.

 \subsection{Elastic wave equation}

 This paragraph details solutions to the elastic wave equation when taking the displacement as  $u_{i}$=$U_{i}e^{-\beta z}exp[i(\omega t-kx)]$. Inserting this expression into Eq. \eqref{eq:EWE} leads to the  system of Eqs. (9,10). Canceling this determinant leads to the following bisquared equation in $q$=$\beta/k$ using the effective elastic tensor coefficients defined in Eq. 12:
 
  \begin{widetext} 
\begin{equation}	q^{4}+\frac{\left(-c_{44}^{2}-c_{11}^{\prime}c_{33}^{\prime}+\left(c_{13}^{\prime}+c_{44}\right)^{2}\right)+\rho V_{r}^{2}\left(c_{33}^{\prime}+c_{44}\right)}{c_{33}^{\prime}c_{44}}q^{2}+\frac{\left(c_{11}^{\prime}-\rho V_{r}^{2}\right)\left(c_{44}-\rho V_{r}^{2}\right)}{c_{33}^{\prime}c_{44}}=0
\end{equation} 
 \end{widetext} 
 
  This equation has two physical solutions, $q_i$=$\beta_{i}/k$ with $U_{z}/U_{x}=r_i$ (i={1,2}) that verify:

\begin{align}	
q_{1}^{2}+q_{2}^{2}&=\frac{\left(c_{44}^{2}+c_{11}^{\prime}c_{33}^{\prime}-\left(c_{13}^{\prime}+c_{44}\right)^{2}\right)-\rho V_{r}^{2}\left(c_{33}^{\prime}+c_{44}\right)}{c_{33}^{\prime}c_{44}}\\
q_{1}^{2}q_{2}^{2}&=\frac{\left(c_{11}^{\prime}-\rho V_{r}^{2}\right)\left(c_{44}-\rho V_{r}^{2}\right)}{c_{33}^{\prime}c_{44}}\\
r_i&=\frac{i\beta_{i}k\left(c_{13}^{\prime}+c_{44}\right)}{k^{2}c_{44}-\beta_{i}^{2}c_{33}^{\prime}-\rho\omega^{2}}(i=1,2)
\end{align}

As neither of these  satisfy the normal boundary condition $\sigma_{xz}|_{z=0}=0$ at the vacuum interface, a linear combinations of these two solutions needs to be considered, as further developed in the text.

 \subsection{Solutions when taking into account the $A_{4\varepsilon}$ term}
 
 At high temperatures ($T\geq T_c/2$), $A_{4\varepsilon}$ (cubic anisotropy) is routinely 10 smaller than  $A_{2\varepsilon}$ (uniaxial anisotropy). At lower temperatures, we rather have $A_{2\varepsilon}\approx$ 4-5 $A_{2\varepsilon}$.  Following the same calculation as in the text but taking into account $A_{4\varepsilon}$ gives the following modified elastic constants:  
 
 \begin{align}
\label{det}
\begin{array}{c}
c_{13}\mapsto c_{13}^{\prime}=c_{13}+A_{\xi}DB\\
c_{11}\mapsto c_{11}^{\prime}=c_{11}-A_{\xi}D^2\\
c_{33}\mapsto c_{33}^{\prime}=c_{33}-A_{\xi}B^2
\end{array}
\end{align}

where the magnetization is aligned by the field along a [$\pm$110] axis and 
 
  \begin{align}
\label{det}
\begin{array}{c}
A_{\xi}=M_{s}sin^{2}(2\theta_{0})\chi_{11}\\
B=A_{2\varepsilon}+\frac{A_{4\varepsilon}}{2}(1+3cos(2\theta_{0}))\\
D=B/2
\end{array}
\end{align}

The shape and position of the resonance remain globally unchanged, but the amplitude of the effect (on both the relative attenuation and the velocity variations) is strongly diminished.

\bibliographystyle{phjcp}

\end{document}